# A Fast and Robust Method for Predicting the Phase Stability of Refractory Complex Concentrated Alloys using Pairwise Mixing Enthalpy

Zhaohan Zhang[a], Mu Li[b], John Cavin[c], Katharine Flores[b,a], Rohan Mishra[b,a,*]

[a]Institute of Materials Science and Engineering, Washington University in St. Louis, St. Louis, MO, 63130, USA

[b]Department of Mechanical Engineering and Materials Science, Washington University in St. Louis, St. Louis, MO, 63130, USA

[c]Department of Physics, Washington University in St. Louis, St. Louis, MO, 63130, USA

## Abstract

The ability to predict the composition- and temperature-dependent stability of refractory complex concentrated alloys (RCCAs) is vital to the design of high-temperature structural alloys. Here, we present a model based on first-principles calculations to predict the thermodynamic stability of multicomponent equimolar solid solutions in a high-throughput manner and apply it to screen over 20,000 compositions. We develop a database that contains pairwise mixing enthalpy of 17 refractory metals using density-functional theory (DFT)-based total energy calculations. To these, we fit thermodynamic solution models that can accurately capture the mixing enthalpy of multicomponent BCC solid solutions. By comparing their energy with DFT-calculated enthalpy of intermetallics from the Materials Project database and using convex hull analyses, we identify the stable phase of any RCCA as a function of temperature. The predicted stability of NbTiZr, NbTiZrV, and NbTiZrV$M$ ($M$ = Mo,Ta,Cr) systems as a function of temperature agree well with prior experimental observations. We apply our model to predict the phase evolution in NbVZr-Ti$_x$ ($0 < x < 1$), which is confirmed experimentally using a high-throughput, laser deposition-based synthesis technique. This method provides a fast and accurate way to estimate the phase stability of new RCCAs to expedite their experimental discovery.

## 1. Introduction

Multi-principal element alloys (MPEAs) are formed by mixing multiple elements at equiatomic or relatively high concentrations. A subset of MPEAs with ≥5 elements that form a single-phase solid solution are called high-entropy alloys (HEAs) [1, 2]. This new alloying strategy has vastly expanded the space of possible alloy systems and led to the discovery of MPEAs with properties not seen in conventional alloys [1, 3, 4]. Despite their compositional complexity, MPEAs tend to form random solid solutions in simple FCC, BCC, or HCP structures, partly due to the increase in configurational entropy with increasing number of elements, which reduces the Gibbs free energy of mixing. However, this "high entropy" effect is insufficient to counteract the driving forces that favor the formation of secondary phases [5]. In fact, among 670 unique MPEAs that were fabricated and characterized between 2004-2020 [6], only 33% of them form a single-phase solid solution or HEAs, while the remaining 67% comprise of coexisting solid solution phases and/or intermetallic phases. As the number of phases and their composition has a direct impact on the mechanical properties of the alloy [7, 8], the ability to rapidly predict the phase stability from the vast compositional space of possible MPEAs is a key step for accelerating their selection and deployment in applications. From a thermodynamic point of view, phase formation in MPEAs is determined by the minimization of total Gibbs free energy ($\Delta G$), including both enthalpy and entropy contributions to the solid solution and any ordered intermetallics that may form for a given alloy composition. The formation enthalpy ($\Delta H_f$) of binary and ternary intermetallics are nowadays readily accessible from computational databases, such as Materials Project, and machine learning models [9-11]; however, quantitatively predicting $\Delta H_f$ of random solid solutions having multiple elements and using them to eventually predict the phase formation in MPEAs remains a challenge.

To aid the selection of HEAs from the broader set of MPEAs, several approaches have been proposed. There are qualitative approaches that involve pairwise analyses of enthalpy and free energy for the constituent binaries within an MPEA to predict the stability of the solid solution [5, 12]. For instance, Troparevsky *et al.* developed a simple criterion to predict which elemental combinations are most likely to form an HEA by setting upper and lower bounds on $\Delta H_f$ of all ordered binary compounds that form for any given combination of elements [13]. Zhang *et al.* proposed a semi-quantitative approach for predicting the mixing enthalpy ($\Delta H_{mix}$) of random solid solutions as the sum of pairwise interactions in the melts of their constituent binaries [14]. Specifically, for a *n*-component MPEA, they expressed $\Delta H_{mix} = \sum_{i=1,i\neq j}^{n} \Omega_{ij} x_i x_j$, where $\Omega_{ij}$ is the regular melt-interaction parameter, which is derived from the mixing enthalpy of $i^{\text{th}}$ and $j^{\text{th}}$ elements in binary liquid alloys using Miedema's model [15], and $x_i$ is the concentration of $i^{\text{th}}$ element. In combination with empirical descriptors, such as the difference in the atomic sizes of the constituent elements, it can separate MPEA compositions that are expected to form solid-solutions from amorphous phases [16]. A more accurate and quantitative method is to directly calculate the total energy of the solid solution using first-principles density-functional theory (DFT) calculations. This requires simulating the random configuration in multicomponent solid solutions using either large supercells of special quasi-random (SQS) structures [17-19] or numerous, small symmetry-inequivalent derivative structures [20, 21]; and calculating either of these using DFT is computationally expensive. Furthermore, the use of SQS to calculate the total energy of individual compositions makes it intractable to handle the large number of possible MPEAs. With regards to the use of derivative structures, Lederer et al. have taken a tour de force approach of using DFT to calculate the total energy of all possible *n*-atom/unit cell ($n \leq 8$) derivative structures for different combination of metals. They then fitted cluster expansion models to these derivatives to parameterize the interaction energies between different atoms and combined it with statistical thermodynamic models to identify temperatures at which a random solid solution is expected to be stable over decomposition into ordered compounds. Their model could correctly predict known solid-solution-forming equimolar binary, ternary, quaternary and quinary alloys, and their crystal structures, *i.e.*, FCC, BCC or HCP with very high accuracy (> 90 %). Despite this remarkable progress, it's rather computationally expensive to parametrize the miscibility-gap and solid solution boundary for a new system. Models that can quickly predict the phase stability for an equimolar MPEA at any given temperature, including the decomposition products — which can include a mixture of solid solution(s) and/or intermetallics — are needed to guide alloy selection and design.

Here, we present a model that can rapidly, and with high accuracy, predict $\Delta H_{mix}$ of equimolar ternary, quaternary, quinary and senary BCC solid solutions that form the class of refractory MPEAs and are popularly referred to as refractory complex concentrated alloys (RCCAs). The alloys involve 17 elements: Ti, V, Cr, Zr, Nb, Mo, Ru, Rh, Hf, Ta, W, Re, Os, Ir, Al, Si, and C. We obtain the $\Delta H_{mix}$ of the multicomponent solid solutions by combining DFT-calculated pairwise mixing enthalpies ($\Delta H_{ij}^{ss}$, where $i$ and $j$ are different elements) of the constituent equimolar binaries with a regular solution model. Our predicted $\Delta H_{mix}$ of 48 equimolar BCC MPEAs show a mean absolute error (MAE) of 16 meV/atom compared to their values calculated directly using DFT with SQS models. We further combine the mixing enthalpy of the solid solutions with the formation enthalpy of intermetallics available in the Materials Project database[22], and use convex hull analyses to predict the most stable phase of ~20,000 equimolar BCC MPEAs at any given temperature with respect to their decomposition products, which can be a mixture of equimolar solid solution(s) and/or intermetallics. We also applied our method to predict the phase evolution in NbTiZr, NbTiZrV, and NbTiZrV$M$ ($M$ = Mo,Ta,Cr) equimolar systems and find the results to be in excellent agreement with experimental observations [7]. Finally, we use our model to predict the phase evolution in NbVZr-$Ti_x$ ($0 < x < 1$) and confirm the results using a high-throughput, laser-processed alloy library. We also note that a similar work was published by Bokas et al. recently [23], where the authors also used pairwise mixing enthalpies to predict the mixing enthalpy of multicomponent solid solutions with high accuracy. Together, these models offer a pathway to accelerate the discovery of multicomponent alloys with desired combination of solid solution and intermetallic phases.

## 2. Methodology

### 2.1. A model for predicting the mixing enthalpy of multicomponent BCC solid solutions

We propose that the pairwise mixing enthalpy, $\Delta H_{ij}^{ss}$, obtained using a regular solution model can adequately describe the interactions in multicomponent solid solutions. We have used a hypothetical equimolar quaternary alloy of elements $A$, $B$, $C$ and $D$ to show the process of calculating its $\Delta H_{mix}$ from pairwise interactions. This process is shown schematically in Fig. 1(a). First, we employ SQS models to generate a series of supercells that approximate a disordered equimolar binary solid solution on a BCC lattice [17]. After testing with different sizes, we select a 24-atom supercell that has a minimal size and shows perfect match to a random alloy when considering 58 pairwise interactions up to the 5th nearest neighbor and 48 triplet interactions up to the 3rd nearest neighbor. Next, we calculate the mixing enthalpies of these binary random solid solutions ($\Delta H_{ij}^{ss}$) with DFT. The computational details are provided in Section 2.3.

With the DFT-calculated $\Delta H_{ij}^{ss}$ of binary random alloys, we can derive the pairwise interaction parameters between a pair of elements using a regular solution model, as shown in Eqn. (1):

$$\Delta H_{ij}^{SS} = \Omega_{ij} x_i x_j, \tag{1}$$

where $i$ and $j$ represent two different elements that form the alloy, $x_i$ and $x_j$ represent their respective concentrations such that $x_i + x_j = 1$, and $\Omega_{ij}$ represents their interaction parameter. From Eqn. (1), we get the pairwise interaction parameter $\Omega_{ij} = \Delta H_{ij}^{ss}/(x_i x_j)$. Then, we estimate the mixing enthalpy ($\Delta H_{mix}$) of the multicomponent BCC solid solutions with a symmetric regular solution model by only considering the interaction energy of the constituent pairs [24], as shown in Eqn. (2) for an *n*-component alloy and Eqn. (3) for the quaternary solid solution *ABCD*:

$$\Delta H_{mix} = \sum_{i=1, i\neq j}^{n} \Omega_{ij} x_i x_j, \tag{2}$$

$$\Delta H_{mix}(ABCD) = \Omega_{AB}.x_A x_B + \Omega_{AC}.x_A x_C + \Omega_{AD}.x_A x_D + \Omega_{BC}.x_B x_C + \Omega_{BD}.x_B x_D + \Omega_{CD}.x_C x_D. \tag{3}$$

As exemplified in Fig. 1(a), for a quaternary alloy *ABCD*, we estimate its mixing enthalpy by summing up $\Omega_{ij} x_i x_j$, where $i, j \in \{A, B, C, D\}$ and $i \neq j$ for all elemental pairs, as shown in Eqn. (3).

With this method, we have calculated $\Delta H_{ij}^{ss}$ of 136 equimolar binary alloys formed by the 17 elements that are highlighted in the Periodic Table in Fig. 1(b) and are frequently used in RCCAs. We plot a heatmap showing $\Delta H_{ij}^{ss}$ between each pair of elements in Fig. 1(c). A green shade indicates a negative value and suggests that the pair of elements favor mixing to form a BCC solid solution compared to their elemental state. A purple shade implies that the pair of elements need extra energy to mix in a BCC phase. A full database of $\Delta H_{ij}^{ss}$ is included in Appendix B.

### 2.2. Predicting the most stable phase(s) of equimolar RCCAs

We determine the most stable phase of RCCAs at any temperature by comparing the thermodynamic stability of competing phases involving all possible equimolar solid solutions and intermetallics. Specifically, for the prediction of RCCAs, which have a BCC crystal structure, we predict $\Delta H_{mix}$ of BCC solid solutions with different number of components using the model discussed above, and retrieve $\Delta H_f$ of binary and ternary intermetallics from the Materials Project database [10]. As most of the elements considered in this work exist in BCC phase, their alloys typically exist in the BCC phase; although there are exceptions [25, 26]. For such exceptional cases, the energy of the solid solution with FCC or HCP phase can also be expressed from the pairwise mixing enthalpy of their binary solutions in their respective FCC or HCP phase, and the phase having the lowest energy could be considered, as has been done recently by Bokas et al. [23].

We next consider the effect of temperature by calculating the Gibbs free energy, $\Delta G_{mix} = \Delta H_{mix} - T\Delta S$. For the BCC solid solutions, we consider the entropy as the ideal configurational entropy of mixing: $\Delta S = \Delta S_{config} = -k_B \sum_{i=1}^{n} x_i ln x_i$, where $k_B$ is the Boltzmann constant, $n$ is the number of elements, $x_i$ is the concentration of $i^{th}$ element and $\sum_{i=1}^{n} x_i = 1$. For the ordered intermetallic phases, $\Delta S$ is set to 0. Other entropy terms, such as vibrational entropy and electronic entropy, which are computationally expensive and are typically one or two orders of magnitude smaller then $T\Delta S_{config}$ [27, 28], have been neglected. Next, we utilize the convex hull analysis as implemented in pymatgen [29] to identify the stability of the BCC solid solution of the RCCAs with respect to decomposition products at any given temperature. The convex-hull construction evaluates the stability of a given phase against any linear combination of compounds that have the same averaged composition [30]. In the quaternary alloy $ABCD$ for instance, we compare its stability with respect to 3 equimolar ternary alloys ($ABC, ABD, BCD$) and 6 binary alloys ($AB, AC, AD, BC, BD, CD$), along with any binary or ternary intermetallics that are reported in the Materials Project database. We define this temperature above which the single-phase solid solution is stable as critical temperature $T_c$ [31, 32]. Below $T_c$, phase decomposition is predicted and at least one element is no longer fully miscible in the solid solution. We apply this analysis to 680 ternary, 2380 quaternary, 6188 quinary, and 12376 senary equimolar refractory HEAs based on the 17 elements included in this study and compare some of the predictions with experimental observations.

We note here that the accurate determination of the thermodynamic phase stability at a given temperature requires not only direct comparison of Gibbs free energies, but also the curvature of Gibbs free energy curve for different phases [33, 34]. Ideally when multiple principal elements mix to form a single-phase BCC solid solution, at a specific temperature, the disordered BCC alloy is stable if it has both lower free energy than any decomposition products, and $\frac{d^2}{dx^2}\Delta G_{mix} > 0$. Curvature determination in high dimensions for multi-principal elements using first-principles calculations requires a dense compositional grid and is computationally intensive. Hence, to rapidly screen stable equimolar RCCAs, we only consider possible decomposition with equimolar sub-alloying compositions, unary metals and stable intermetallics as reactants. The solid solution phase is identified to be stable when it exhibits the lowest energy against the above decomposition reactions. For a specific alloy system of interest, we can employ our approach with dense composition grids taking non-equimolar alloys into account to determine its phase diagram[35], as we have discussed later in the Results section.

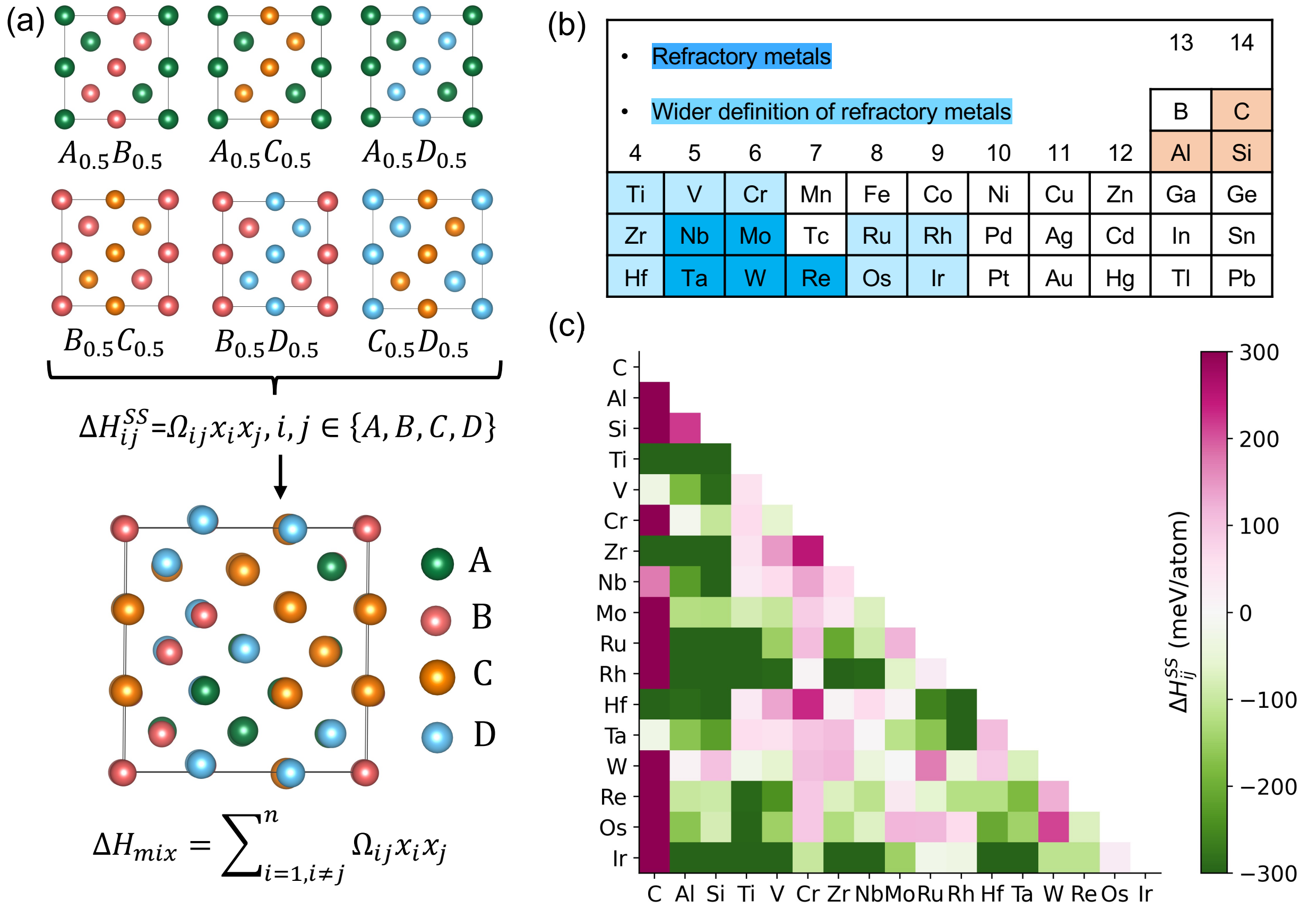

Fig. 1. (a) Schematic showing the approach we have used to calculate the enthalpy of mixing of a quaternary equimolar BCC solid solution of elements *A, B, C,* and *D* using their pairwise mixing enthalpy. (b) We calculate binary interaction parameters, $\Omega_{ij}$, of 17 elements, including 5 refractory metals, 9 quasi-refractory metals and Al, C, and Si. (c) Heatmap showing the pairwise mixing enthalpy (in units of meV/atom) for the 136 pairs formed of the 17 elements.

### 2.3. Computational details

To model the disordered BCC solid solution, we generated SQSs [36] using the Alloy Theoretic Automated Toolkit (ATAT) [37]. For the high-throughput calculation of pairwise mixing enthalpy, we used a 24-atom supercell. For the simulation of ternary, quarternary and quinary alloys, we selected 36-atom, 48-atom, and 60-atom supercells, respectively [38]. We set the range of pair clusters to the 5$^{th}$ nearest neighbor and the triplets to the 3$^{rd}$ nearest neighbor. We performed first-principles DFT calculations using the Vienna Ab initio Simulation Package (VASP) [39]. We employed the generalized gradient approximation (GGA) as implemented in the Perdew-Burke-Ernzerhof (PBE) [40] exchange-correlation functional to approximate the many-body electronic interactions. We used the projector augmented-wave (PAW) method [41] to describe the core electrons with the outer $p$ semicore states included as valence states. For the calculation of pairwise mixing enthalpy $\left(\Delta H_{ij}^{ss}\right)$, we fixed the plane-wave energy cutoff at 520 eV and relaxed the structures until the forces on each atom were less than 0.001 eV Å$^{-1}$. For the 24-atom supercell, a grid density of 10000 k-points/number of atoms was used. The binary $\Delta H_{ij}^{ss}$ is calculated with respect to constituent elements in their stable states ($E(i)$), as shown in Eqn. (4). $\Delta H_{mix}$ in multicomponent solid solutions were calculated in a similar manner. An example of $\Delta H_{mix}$ for a quaternary alloy is shown in Eqn. (5):

$$\Delta H_{ij}^{SS}(AB) = E(A_{0.5}B_{0.5}) - 0.5E_{el}(A) - 0.5E_{el}(B). \quad (4)$$

$$\Delta \mathrm{H}_{\mathrm{mix}}(ABCD) = E(A_{0.25}B_{0.25}C_{0.25}D_{0.25}) - 0.25E_{el}(\mathrm{A}) - 0.25E_{el}(\mathrm{B}) - 0.25E_{el}(\mathrm{C}) - 0.25E_{el}(\mathrm{D}). \quad (5)$$

### 2.4. Deposition of NbVZr-Ti$_x$ alloy library and its characterization

A 20 × 20 × 5 mm equiatomic NbVZr substrate was produced using arc melting and casting. The raw materials with purity ≥ 99.8 wt. % were melted on a water-cooled copper hearth in an argon atmosphere. The obtained buttons were flipped and remelted at least five times to improve compositional homogeneity prior to casing into a copper mold to produce a 20 × 20 × 5 mm plate. The composition libraries were prepared using an Optomec MR-7 Laser Engineered Net Shaping system (LENS$^{TM}$). On the NbVZr substrate, a total of 16 patches, each sized 2 mm × 2 mm, were alloyed by injecting varying amount of Ti powder (-100+325 mesh, ⩾ 99.5 % purity) into a melt pool created by the moving laser to produce NbVZrTi$_x$ alloys ($x$ = 0 – 1). The powder feed rates varied from 2.0 rpm to 3.5 rpm in increments of 0.1 rpm, while the laser power and travel speed were held constant at 250 W and 6.35 mm/s, respectively. Each single-layered patch consisted of 5 parallel laser tracks, and there was approximately 25% overlap between adjacent tracks. Subsequently, the patches in the library were remelted twice with a 250 W laser to ensure proper mixing between Ti and the substrate material. More detailed description of the library preparation can be found elsewhere [42].

The crystal structures were characterized on the polished library surface (plan view) using X-ray diffraction (XRD, Rigaku D-Max/A) with Cu-K$_\alpha$ energy. Diffraction angles between 20° and 100° (2θ) were collected with a step size of 0.02°. To eliminate the effect of the surrounding materials and isolate the patches with varying Ti content, a plexiglass mask with a tapered circular aperture (2 mm in diameter) was employed. The diffraction signal from the mask was subtracted from the overall XRD patterns. The microstructures of the patches were characterized using a field emission scanning electron microscope (SEM, JEOL JSM-7001FLV) operated at 15 kV accelerating voltage. Their composition was evaluated using an Oxford Aztec Live X-Max energy dispersive X-ray spectroscopic system (EDS).

## 3. Results

### 3.1. Prediction accuracy of $\Delta H_{mix}$

To verify the accuracy of our method, we compare the model prediction to the $\Delta H_{mix}$ of 69 equimolar refractory HEAs with DFT-calculated values obtained using SQS models. Among the 69 investigated compositions, 37 of them are experimentally reported alloys that we collected from the literature [6, 43], while the rest are generated by randomly selecting combinations of the refractory metals used in this study. For each alloy, we plot the error in the model-predicted $\Delta H_{mix}$ with respect to the DFT-calculated value in the top panel for ternary compositions in Fig. 2(a), quaternary compositions in Fig. 2(b), and quinary compositions in Fig. 2(c). The alloy compositions are labeled at the bottom of each figure. The areas shaded yellow, blue and violet represent, respectively, 7 ternary, 15 quaternary and 15 quinary compositions that have been reported in literature [6, 43]. The compositions in the unshaded areas have been generated randomly. The mean absolute error (MAE) for ternary, quaternary and quinary alloys are 15, 17, and 22 meV/atom, respectively. The small MAE values demonstrate that our model shows good accuracy in predicting the $\Delta H_{mix}$. 80% of the prediction falls within the error range between -25 to 25 meV/atom, as indicated by the area between the two horizontal green lines. For reference, the MAE of DFT-calculated $\Delta H_f$ with respect to experimental measurements is ~ 0.145 eV/atom for entries in the Materials Project database, when using the elemental DFT total energies as chemical potentials [44, 45].

In addition, we use the same alloy compositions to test the accuracy of the regular-melt model developed by Zhang et al. [14], which are shown in the middle panel in Figs. 2(a-c). In the regular-melt model, the mixing enthalpies of binary liquid alloys from Miedema's model ($\Delta H_{ij}^{liq}$) [15] are used as fitting parameters. The MAE for ternary, quaternary and quinary compositions are 82, 83, and 83 meV/atom, respectively. We also repeat the analysis using the lowest formation enthalpy of stable binary intermetallics ($\Delta H_{ij}^{int}$) that were collected by Troparevskey et al. from the AFLOW database[13, 46], which results in MAE of 116, 123, and 189 meV/atom for ternary, quaternary and quinary compositions, respectively. Thus, we find that combining DFT calculations done on binary alloys with a regular solution model gives accurate prediction of $\Delta H_{mix}$ in multicomponent solid solutions. The pairwise interactions obtained from a disordered BCC lattice, as opposed to liquid alloys or ordered intermetallics, are necessary for making accurate predictions. The simulation of binary solid solutions requires small 24-atom supercells and are less computationally intensive than directly modeling the multicomponent solid solutions, which will require (36-120)-atom supercell for ternary, quaternary, and quinary systems, not to mention the vast combinatorial composition space of those alloys, each of which will require a separate SQS. By constructing a binary $\Delta H_{ij}^{SS}$ database, we can efficiently screen a large number of multicomponent systems and assist the design of RCCAs. In this work, with only 136 computations on pairs formed by the 17 elements, we predict $\Delta H_{mix}$ of over 20,000 equimolar multicomponent compositions. We have made this repository available online [38].

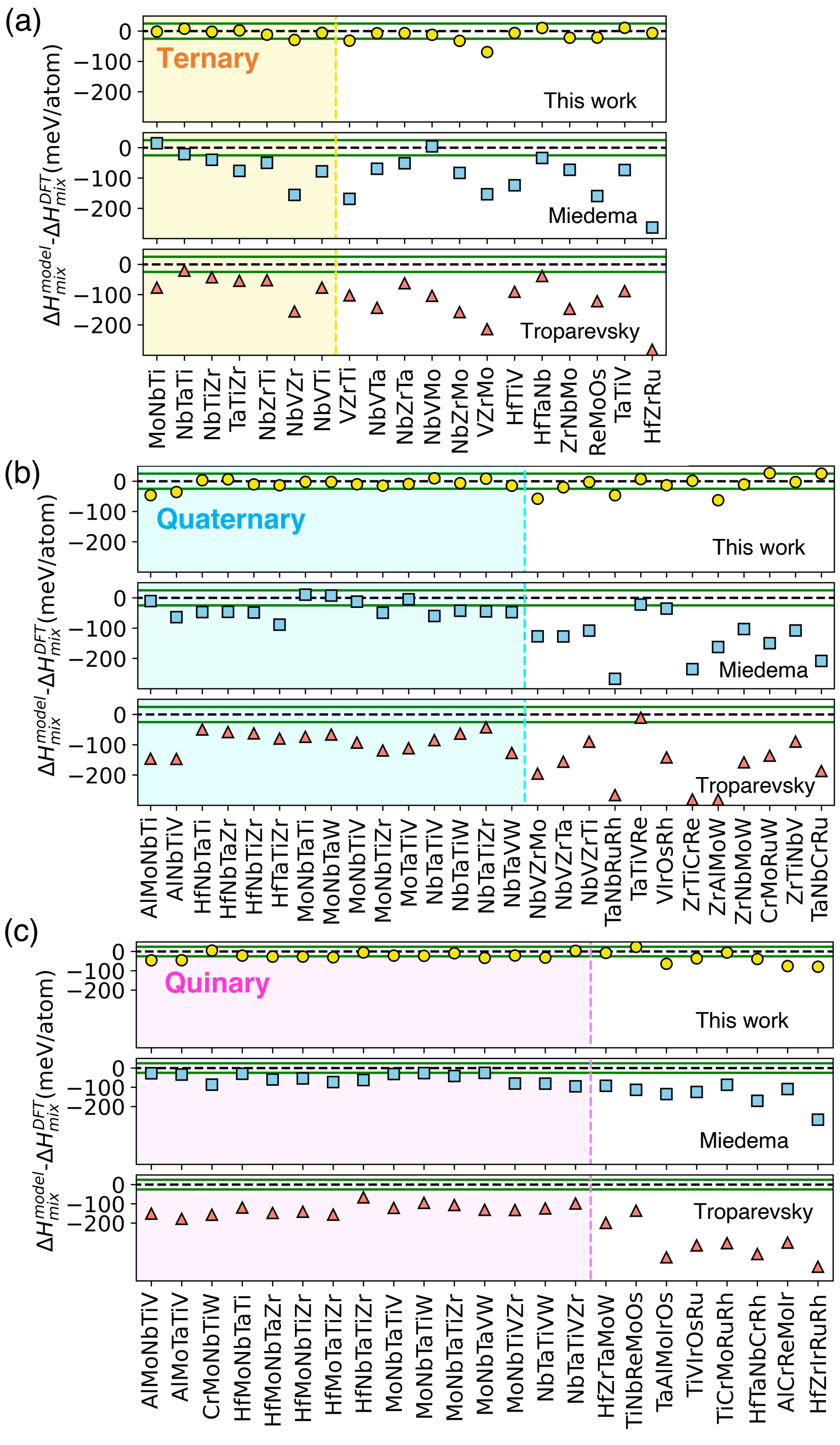

Fig. 2. A comparison of $\Delta H_{mix}$ predicted using regular solution models with their DFT-calculated value for (a) ternary, (b) quaternary, and (c) quinary equimolar RCCAs. For each alloy family, the top panel represents the predictions from this work, where the interaction parameters for the regular solution model were obtained using DFT-calculated pairwise mixing enthalpies ($\Delta H_{ij}^{SS}$). The middle

panel shows results obtained using the mixing enthalpy of binary liquid alloys from Miedema's model ($\Delta H_{ij}^{liq}$)[15]. The bottom panel uses the formation enthalpy of binary intermetallics collected by Troparevsky et al. ($\Delta H_{ij}^{int}$)[13]. The *y*-axis shows the prediction error in units of meV/atom for each model. The two horizontal green lines in each panel indicate the range of errors between -25 to 25 meV/atom. The areas shaded yellow, blue and violet represent, respectively, 7 ternary, 15 quaternary and 15 quinary compositions that have been reported in literature [6, 43]. The compositions in the unshaded areas have been generated randomly.

## 3.2 Phase predictions validated by experiments

### 3.2.1 NbTiZr-based equimolar refractory HEAs

Having established that pairwise mixing enthalpies, $\Delta H_{ij}^{SS}$, can predict $\Delta H_{mix}$ of multicomponent solid solutions with similar accuracy as DFT, we next compare our model prediction of phase evolution as a function of temperature for specific RCCAs where experimental observations are available. Here, we use NbTiZr-based RCCAs as a model system to demonstrate the convex hull analysis process and benchmark our predictions with experimental observations from Senkov et al. [7]. Starting with the ternary alloy NbTiZr, we first retrieve $\Delta H_{ij}^{SS}$ of the 3 constituent binary compositions NbTi, NbZr, and TiZr from our database, and use them to predict $\Delta H_{mix}$ of NbTiZr. We then query all binary and ternary entries that are reported in the Nb-Ti-Zr chemical space in the Materials Project database. In this case, we obtain 11 binary ordered compounds and 1 ternary compound, whose $\Delta H_f$ are all positive and range between 34 – 155 meV/atom, as shown in Table 2 in Appendix C. The $\Delta H_f$ of the three constituent elements, Nb, Ti, Zr, is set to 0 following Eqn. (4). We then calculate $\Delta G$ of all 19 phases by assuming ideal configurational entropy of mixing and conduct the convex hull analyses for temperatures ranging from 0 to 3000 K, in steps of 200 K. At each temperature, we assess the stability of a given phase versus its decomposition to any linear combination of possible phases that give the same average composition. A convex hull is determined by combining all stable phase points such that any linear combination of possible phases lies on or above the convex hull, i.e., have the same or higher $\Delta G$. On the convex hull boundary, the energy curve is convex such that $\frac{d^2}{dx^2}\Delta G_{mix} \geq 0$. Therefore, we represent the stability of a phase by the term energy above hull ($E_{hull}$). If a phase is stable (i.e. on the convex hull), $E_{hull} = 0$; if not (i.e., it is above the convex hull), $E_{hull}$ is positive and it will decompose into phases with lower energy. We plot $E_{hull}$ as a function of temperature for NbTiZr alloy in Fig. 3(a). The regions shaded with different colors represent the ranges of temperatures wherein different combination of phases are the most stable, and the unshaded region shows the temperature range where the multicomponent BCC solid solution is most stable. At 0 K, we find that NbTiZr BCC solid solution is unstable with $E_{hull}$ = 25 meV/atom. It is expected to separate into elemental Nb, Ti, and Zr. With increasing temperature, the increasing contribution of $-T\Delta S_{mix}$ reduces the Gibbs free energy of the ternary solid solution, as indicated by the yellow curve in Fig. 3(a). For temperatures > 1000 K, the NbTiZr solid solution becomes stable as it lies on the convex hull. Our results agree well with experimental observations after annealing at 1400 °C for 6 hours [7], where the alloying elements were observed to be homogeneously distributed in equiaxed BCC grains. We define this temperature above which the single-phase solid solution is stable as critical temperature $T_c$. We note that the temperature step we use here is 200 K, so $T_c$ can have an error range $\pm$200K. Other neglected entropy terms, such as vibrational entropy, will also impact $T_c$.

Next, we investigate the phase stability in the quaternary system with the addition of V to NbTiZr. For the analysis of NbTiZrV, we construct the convex hull with 1 quaternary solid solution, 4 ternary solid solutions, 6 binary solid solutions, 4 unary metals, as well as the intermetallic entries, as shown in Table 3 in Appendix C. We predict the quaternary alloy to stabilize as a single-phase BCC solid solution, i.e., $E_{hull}$ = 0, above 1400 K, as indicated by the green curve in Fig. 3(b). At lower temperatures, the most stable phases comprise of a NbTiZr-rich BCC phase and a $NbV_2$-rich Laves phase. Experimentally, NbTiZrV is found to have a dominant BCC phases with clusters of fine, V-rich precipitates(22Nb-21Ti-18Zr-39V at.%) inside NbTiZrV grains [7], which is consistent with our prediction.

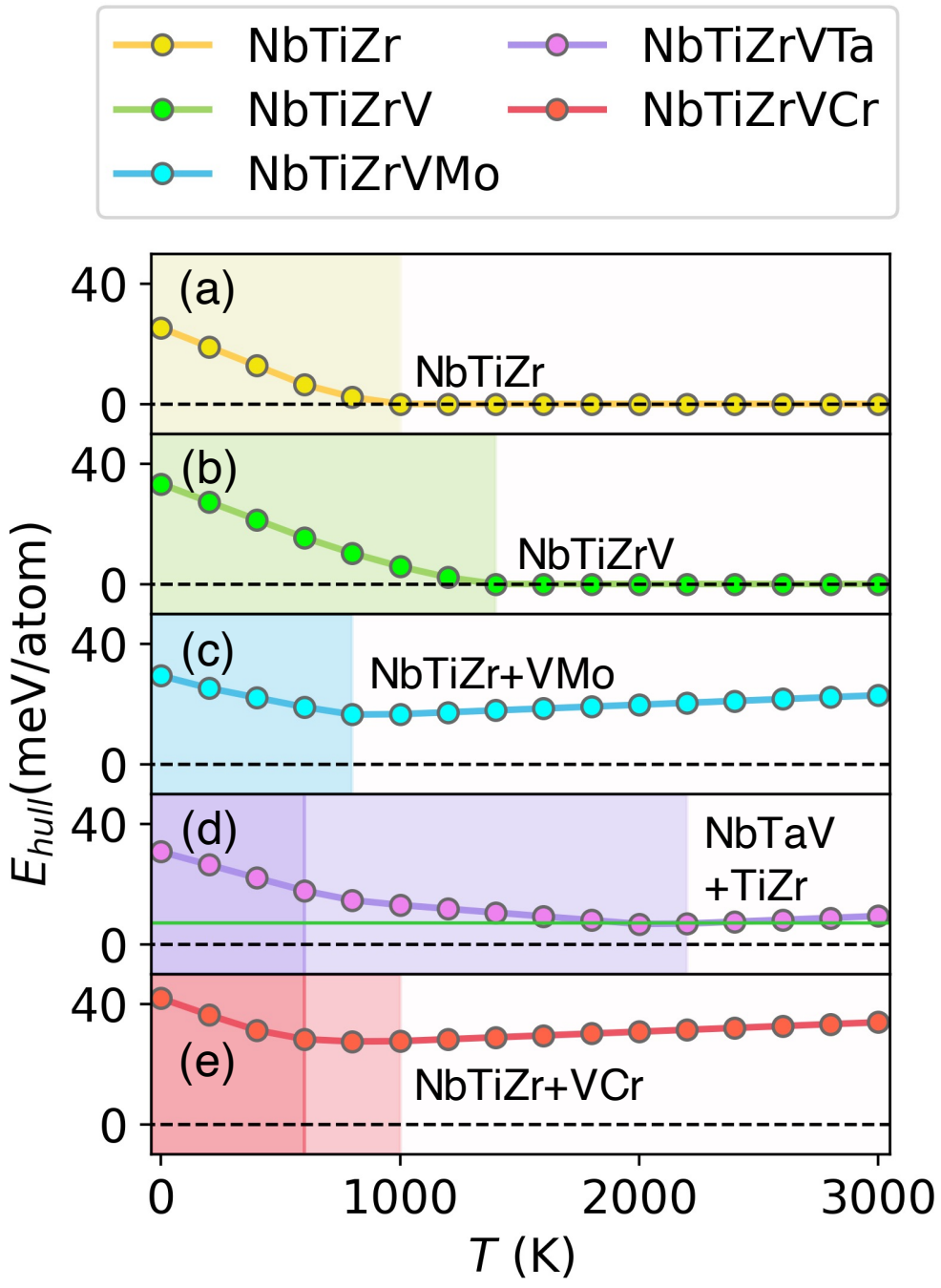

Fig. 3: Predicted phase stability of NbTiZr-based refractory HEAs. (a) NbTiZr and (b) NbTiZrV single-phase BCC solid solutions become stable at 1000K and 1400K, respectively; (c) NbTiZrVMo, (d) NbTiZrVTa, and (e) NbTiZrVCr are predicted to stabilize as two BCC phases at higher temperatures. The shaded areas in each panel

indicate the temperatures at which the ground-state phases include one or more intermetallics.

Further alloying of NbTiZrV with a fifth element, Mo, Ta or Cr introduces more possible phases that range from 5-component compositions to pure elements. For NbTiZrVMo, when the temperature ranges from (0 – 800) K, the most stable phases consist of MoTi-rich BCC, $NbV_2$-rich Laves and Zr-segregated phases, as shown by the region shaded in blue in Fig. 3(c). For temperatures >800 K, we predict the alloy to stabilize as two separate BCC phases, NbTiZr and MoV. In experiments, after homogenization annealing at 1400 ℃, NbTiZrVMo shows a dendritic microstructure consisting of three phases: Zr-depleted BCC-1, Zr-rich BCC-2, and $(Mo, V)_2Zr$ Laves phase [7]. Hence our thermodynamic model successfully predicts the two BCC phases at higher temperatures, while the Laves phase is captured at lower temperatures. The observation of Laves precipitates at high temperatures in experiments could be due to other thermodynamic factors, such as interface energy between the precipitates and the matrix, or kinetic factors such as the sluggish diffusion observed in many HEAs[47]. Our prediction agrees well with a first-principles study on NbTiZrVMo [48], where the authors investigated the short-range ordering in Nb-Ti-Zr-V-Mo system using Monte Carlo simulations, and also observed that Mo and V tend to cluster while Zr tends to separate from Mo and V.

We conducted a similar analysis on NbTiZrVTa and predict several phase transitions with temperature, as shown in Fig. 3(d). Each shaded area indicates a region where a different phase or combination of phases is most stable. From left to right, for temperature < 600 K, the stable mixture consists of a solid solution of NbTa that we call BCC1, a solid solution of NbTi or BCC2, $TaV_2$ Laves, and a Zr-segregated phase; for temperatures between (600 – 2200) K, the most stable phases are NbTiZr-BCC1, NbTa-BCC2, TiZr-BCC3, and $TaV_2$ Laves phases; and above 2200 K, the Laves phase is suppressed and the alloy stabilizes into NbTaV-BCC1 and TiZr-BCC2. For reference, the average melting temperature of these five elements is 2462 K. Experimentally, NbTiZrVTa consists of two BCC phases in hot-worked conditions and is a single-phase BCC solid solution after annealing at 1400 ℃ [7]. This discrepancy between our predictions and the experimental microstructure after annealing could be due to the very small value of $E_{hull} = 7$ meV/atom for the NbTiZrVTa BCC phase predicted using our model, which is below the MAE of 22 meV/atom for quinary systems. Furthermore, a positive energy is required for the nucleation of a secondary phase and the formation of interfaces between the product phases. Therefore, phases with small $E_{hull}$ values, can be expected to be stable.

We notice similar phase transition trends in NbTiZrVCr. At low temperatures ranging from (0 – 400) K, it shows NbVCr clustering and segregation of Zr and Ti. From 600 to 1000 K, configurational entropy favors the mixing of NbTi and VCr, while Zr remains segregated. When the temperature goes above 1000 K, it stabilizes into NbTiZr-BCC1 and VCr-BCC2, as shown in Fig. 3(e). Experimentally, the annealed sample comprises of (Nb, Ti)-rich BCC1 and (V, Cr)-rich Laves, with Zr distributed in both the phases. Comparing the experimental microstructure with our prediction, we successfully capture the clustering tendency of Nb-Ti and V-Cr. However, the exact Laves phases were missed since we ignored the entropy stabilization effect at higher temperature for intermetallic phases, which will be discussed in Section 4.

### 3.2.2 Phase evolution in NbVZr-Ti$_x$ ($0 < x < 1$) alloys

We further apply our method to assist the design of microstructure in RCCAs using NbVZr as a base alloy. Equimolar NbVZr exhibits a dendritic microstructure consisting of a BCC solid solution, with two Laves structures forming in the interdendritic regions [43]. The secondary Laves phases have been reported to strengthen the alloy by acting as obstacles to dislocation motion. To investigate the effect of composition on the microstructure evolution in the NbVZr-base alloy system, we alloy it with Ti and predict the phase stability as a function of Ti concentration [49]. We first predict the energy of non-equimolar NbVZr-Ti$_x$ ($0 < x < 1$) alloys using a regular solution model, as shown in Eqn. (3), where $x_{\mathrm{Nb}} = x_{\mathrm{V}} = x_{\mathrm{Zr}} = \frac{1}{3+x}$, $x_{\mathrm{Ti}} = \frac{x}{3+x}$, and $x$ is in the range of $[0, 1]$ with a grid spacing of 0.1. For every non-equimolar NbVZr-Ti$_x$ composition, we conduct convex hull analysis and plot $E_{hull}$, in the units of meV/atom, of the NbVZr-Ti$_x$ BCC phase as a function of $x$ and temperature as a heatmap in Fig. 4(a). At 0 K, the quarternary alloy tends to decompose into $NbV_2$-rich Laves, and segregated Zr, Nb, and Ti phases. With increasing temperature, NbZrTi-BCC phase is then stabilized by configurational entropy, with the precipitation of $NbV_2$ Laves phase. Finally, the quaternary NbVZrTi$_x$ BCC phase becomes stable at elevated temperatures. The black dashed line in the heatmap demarcates the composition and temperature at which a single-phase BCC alloy becomes stable, i.e., its $E_{hull} = 0$. Below the dashed line, it forms as a multi-phase alloy that contains BCC and Laves phases. For the ternary NbVZr alloy, the critical temperature $T_c$, at which their BCC solid solution becomes stable, is quite high at 2400 K; with increasing Ti concentration, $T_c$ decreases, which suggests that NbVZr-Ti$_x$ alloys would tend to stabilize as BCC solid solution with increasing Ti fraction.

To confirm this prediction, we fabricated a laser-processed NbVZr-Ti$_x$ library and characterized their crystal structures as a function of composition with X-ray diffraction. The results are as shown in Fig. 4(b), from which we can see that Ti addition reduces the Laves phase fraction and results in a single-phase BCC solid solution in equimolar NbVZrTi.

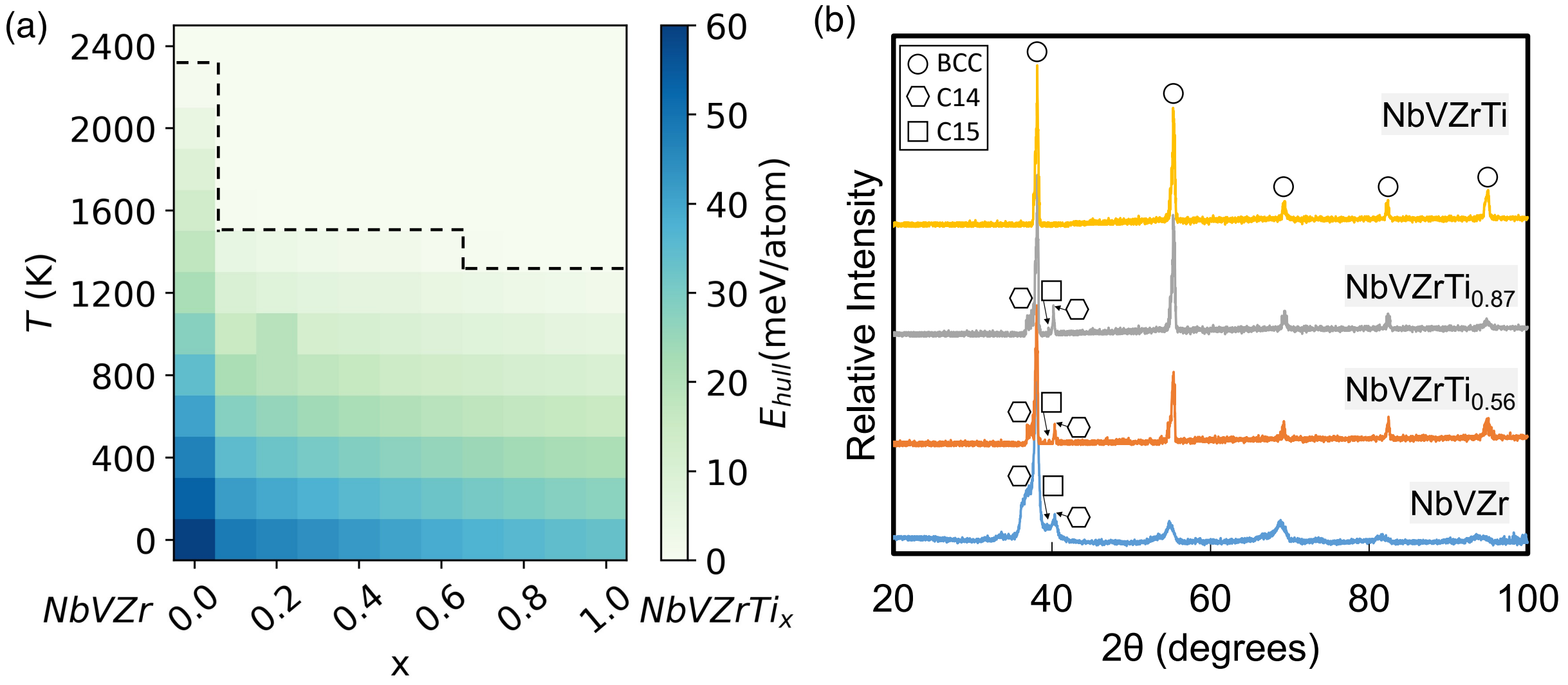

Fig. 4: Phase stability of NbVZr-based refractory HEAs: (a) Model-predicted BCC phase stability in NbVZrTi$_x$ ($0 < x < 1$) alloys. The colormap indicates $E_{hull}$ of NbVZrTi$_x$ as a function of Ti concentration, $x$, and temperature. When $E_{hull} = 0$, a single-phase BCC solid solution is stable. (b) X-ray diffraction of NbVZrTi$_x$ library. Ti addition eliminates the C14 and C15 Laves phases and results in a single BCC phase at equimolar composition.

### 3.3 Candidate RCCAs systems

With the above examples, we have shown that our model can predict the experimental phase-stability of both equimolar and non-equimolar RCCAs. We next apply it to identify new RCCA systems for future research. Here, based on the 17 elements in our dataset, we investigate the stability 680 ternary, 2380 quaternary, 6188 quinary, 12376 senary equimolar compositions. For each of them, we perform convex hull analyses comparing the energy of multicomponent BCC phase with all the lower order equimolar solid solutions and binary/ternary intermetallic phases retrieved from the Materials Project database. Using $E_{hull} = 0$ meV/atom as a standard for a stable phase, we plot the number of single-phase BCC RCCAs that are predicted at 1000 K and 2000 K, respectively, in the top panel of Fig. 5(a). We observe an overall reduced number of single-phase RCCAs with increasing number of elements due to the competing solid solutions and intermetallic phases introduced by the constituent elements. More single-phase RCCAs are stabilized at 2000 K compared to 1000 K as a result of the increase in configurational entropy at higher temperatures. Considering the energy required for the nucleation of secondary phases and the formation of interfaces during phase decomposition, we benchmark the metastability of the multicomponent BCC phase as $E_{hull} \leq 20$ meV/atom. This value is selected after comparing the $E_{hull}$ of experimentally known single-phase RCCAs, as listed in Table 1 and shown in Fig. 6 in Appendix B. With this criterion, the number of synthesizable single-phase RCCAs increases, as shown in the bottom panel of Fig. 5(a).

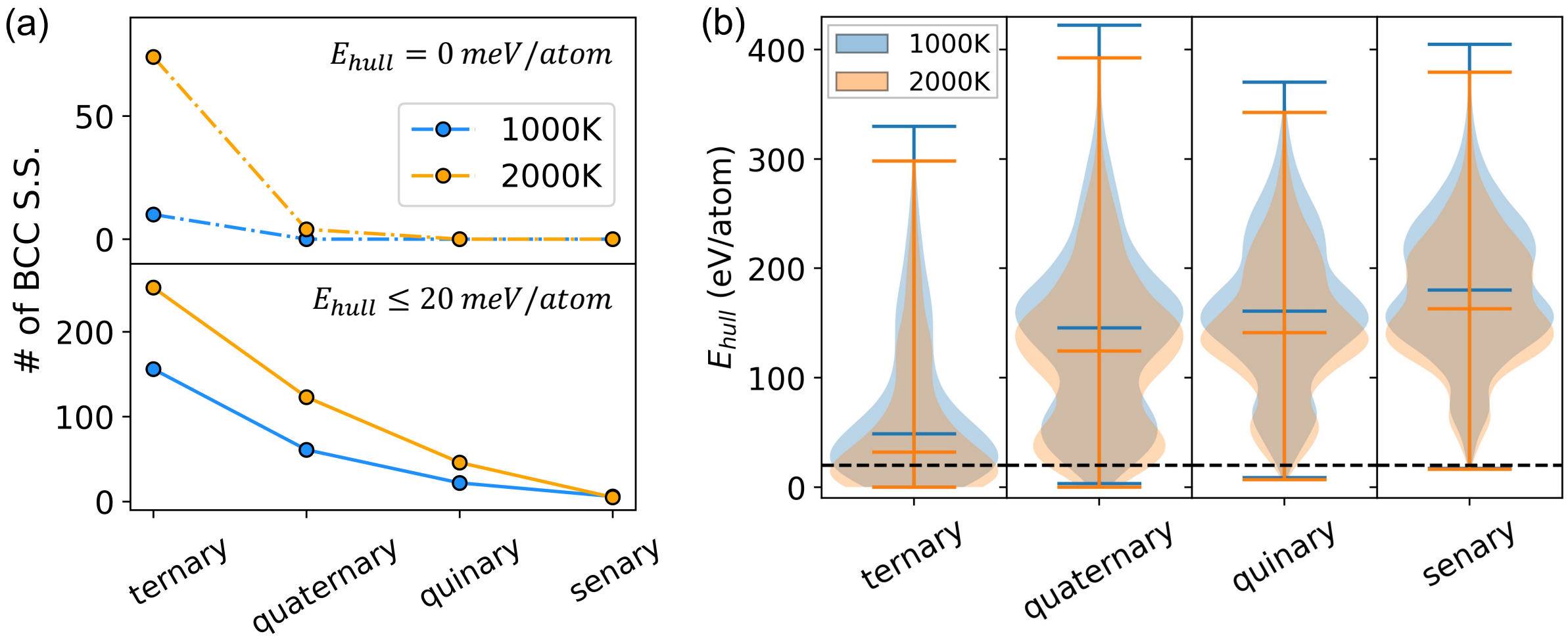

Fig. 5: (a) Number of RCCAs predicted to stabilize as a single-phase BCC solid solution out of the 680 ternary, 2380 quaternary, 6188 quinary, and 12376 senary equimolar alloys formed using the 17 elements used in this study. The results in the top panel were obtained with a stability criterion of $E_{hull}$ = 0 meV/atom, and the ones in bottom panel were obtained using $E_{hull}$ = 20 meV/atom. (b) The distribution of the predicted $E_{hull}$ values for the BCC phase of all the multicomponent equimolar alloys in each of the four categories.

To visualize the relationship between the number of elements in RCCAs and their stability, we plot the distribution of $E_{hull}$ for over 20,000 compositions in the category of ternary, quarternary, quinary and senary alloy systems, as shown in Fig. 5(b). In these violin plots, the lower, middle, and upper bars represent the minimum, median, and max $E_{hull}$ values, respectively. We mark $E_{hull}$ = 20 meV/atom with the black dashed line as the value for formability. The blue violin indicates predictions at 1000 K and the orange violin is for predictions at 2000 K. In each box, the orange violin shifts to lower $E_{hull}$ values, signifying stabilization due to the increase in configurational entropy at higher temperatures. From left to right, we detect a shift of $E_{hull}$ from lower to higher values. The fraction of compositions with $E_{hull} \leq 20$ meV/atom in ternary, quarternary, quinary, senary systems are 38.1%, 5.4%, 0.8%, and 0.1%, respectively, which shows that alloying with more elements does not necessarily lead to a single-phase solid solution. The formation of single-phase RCCAs requires a favorable mixing enthalpy of the constituent elements, as well as absence of a strong tendency for clustering — that can drive the formation of lower order solid solutions or intermetallics. With increasing number of constituent elements, the competing phases become more complex. In fact, even though the combinatorial chemical space is enormous with more elements, the percentage of alloys that favor formation of a single-phase drops dramatically, as shown in Fig. 5(a). We list the compositions that are predicted with $E_{hull} \leq 20$ meV/atom at 1000 K and 2000 K in the Appendix D. We have also shared the entire prediction dataset and example scripts to construct convex hull for specific alloy systems on a publicly available repository [38].

## 4. Discussion

The model we have developed can predict the stable phases and decomposition energies of a RCCA by using convex hull analyses. We show that it can successfully capture phase evolution observed experimentally in many refractory HEA systems. However, the current model does not include certain effects that may result in different outcomes. We list some of those factors below:

One factor is the value of $E_{hull}$ below which a single-phase solid solution can be stabilized in experiments. When $E_{hull} = 0$ at a given temperature, the solid solution becomes the lowest energy phase. For positive values of $E_{hull}$, the system can lower its energy by decomposing into other stable phases. Yet, additional energy is needed for atoms to diffuse and for the nucleation of the secondary phase along with the formation of new interfaces. If the decomposition energy of the higher-order solid solution is not sufficient for the nucleation events, it is expected to be stable even though $E_{hull}$ is positive. The energy scale for these complex processes can be estimated by comparing the $E_{hull}$ of known single-phase RCCAs, as has been done for inorganic crystalline compounds by *Sun et al.* [50]. We have used the extremely limited set of 4 ternary, 15 quaternary and 15 quinary RCCAs, as listed in Table 1 in Appendix B, which have been characterized to form single-phase BCC alloy in experiments [6], and fit a Gaussian kernel density to their DFT-calculated $E_{hull}$ value at 0 K. The resulting probability distribution for single-phase ternary, quaternary and quinary RCCAs as a function of their $E_{hull}$ value is shown in Fig. 6 in Appendix B. While ternary and quaternary compositions show a metastability up to 20 meV/atom, quinary compositions exhibit higher metastability up to 50 meV/atom.

A second factor is the level of lower-order alloy systems included in the convex hull analyses. For the results presented above, we use the ground state energy of the constituent phases to identify thermodynamic equilibrium. That is, for an *N*-component system, we involve equimolar solid solution phases ranging from *N*-component, (*N* – 1)-component, …, to binary and unary, along with the intermetallic entries to determine the phase stability of equimolar RCCAs. We can instead use metastable decomposition products that may be observed in experiments due to limited interdiffusion, for example. This is especially relevant for certain combinatorial approaches for rapid alloy development [51], such as additive manufacturing. As an example, we compare the results of all-inclusive stability analysis with an analysis that only considers $n$-component versus ($n$-1)-component solid solution phases for 27 experimental quaternary alloys. All intermetallic phases are included in both the cases. The all-inclusive stability analysis predicts 10 of them to form single-phase solid solution with $E_{hull} = 0$, while the analysis that only considers decomposition into ternary solid solutions and intermetallic phases predicts 18 of them with $E_{hull} = 0$. Therefore, one can amend this parameter for the specific alloy fabrication method.

Lastly, we discuss the role of various entropy factors that were not included in our model. We assume that in the solid solution phases, the elements are randomly distributed and use the ideal configurational entropy of mixing for a fast assessment. This simplistic treatment is likely to miss effects such as short-range ordering that are frequently observed in experiments [8, 52]. This can be improved by using sub-regular solution models of mixing enthalpy that are fairly scalable [53, 54]. For greater accuracy, at the expense of speed, first-principles-based methods combined with molecular dynamics or Monte Carlo simulations can be used [55, 56]. We have also left out the entropy contributions in intermetallic phases by assuming that they are fully ordered. However, intermetallic phases in RCCAs usually have more constituents than sub-lattices, so that two or more elements will usually occupy one sub-lattice[57, 58]. For instance, consider an $L1_2$ phase in a quaternary alloy (*ABCDE*), if the intermetallic phase has a random distribution of elements *A* and *B* on one sub-lattice and a random distribution of elements *C, D* and *E* on the second sub-lattice, its configurational entropy will be equal to over 60% of the configurational entropy of the equimolar BCC solid solution *ABCDE*. In the future, we will include complex intermetallic phases and both the enthalpy and entropy contribution to refine our model. Moreover, we have excluded the contribution of vibrational enthalpy and entropy to the free energy of both the solid solution and the intermetallic phases. While the absolute magnitude of vibrational entropy ($S_{vib}$) can be much larger than configurational entropy, especially, at elevated temperatures [59, 60], the difference in vibrational entropy ($\Delta S_{vib}$) between two phases having the same composition or between the parent and product phases, is often smaller than $\Delta S_{config}$ [61]. For instance, the vibrational free energies of the BCC and the Laves phases are nearly equal in CrMoNbV RCCAs [18]. Another study on a large set of binary alloys reported that the contribution of vibrational entropy to free energy is much smaller than that of configurational entropy for the majority of alloys studied in that work; however, in certain alloys with shallow mixing enthalpy, vibrational entropy had to be considered to accurately predict their miscibility. Thus, going forward, efficient approaches to calculate the vibrational free energy from first-principles calculations [62], followed by their parameterization, can make our model more accurate.

## 5. Conclusion

In summary, we present a fast and accurate thermodynamic method to predict the phase stability of RCCAs in a high-throughput manner. We show that pairwise mixing enthalpy is enough to give accurate prediction to the mixing enthalpy of multi-component solid solutions. Further, with convex hull analysis, we can construct phase diagrams of alloy systems that agree well with experimental observations. Finally, we screen over 20,000 RCCAs and investigate their thermodynamic stability at different temperatures. We propose our model to be a convenient tool to predict the phase stability of RCCAs and aid their experimental discovery. A python code to construct our model and the predicted datasets are available online [38].

**Acknowledgements**

This work was primarily supported by the National Science Foundation (NSF) under Grant No. DMR-1809571. J.C. and R.M. acknowledge support through NSF Grant No. CBET-1729787. This work used computational resources of the Extreme Science and Engineering Discovery Environment (XSEDE), which is supported by NSF through Grant No. ACI-1548562. The authors acknowledge financial support from Washington University in St. Louis and the Institute of Materials Science and Engineering for the use of shared instruments and staff assistance.

**Appendix A. Pairwise mixing enthalpy database (unit: meV/atom)**

| | Mo | Nb | Ta | V | W | Zr | Ti | Al | Hf | Cr | C | Re | Ru | Os | Rh | Ir | Si |
|---|---|---|---|---|---|---|---|---|---|---|---|---|---|---|---|---|---|
| Mo | 0 | | | | | | | | | | | | | | | | |
| Nb | -73 | 0 | | | | | | | | | | | | | | | |
| Ta | -116 | 0 | 0 | | | | | | | | | | | | | | |
| V | -102 | 65 | 55 | 0 | | | | | | | | | | | | | |
| W | -2 | -41 | -76 | -59 | 0 | | | | | | | | | | | | |
| Zr | 42 | 64 | 109 | 146 | 119 | 0 | | | | | | | | | | | |
| Ti | -81 | 35 | 61 | 52 | -21 | 49 | 0 | | | | | | | | | | |
| Al | -123 | -223 | -165 | -183 | 14 | -333 | -307 | 0 | | | | | | | | | |
| Hf | 11 | 63 | 109 | 136 | 90 | 9 | 36 | -290 | 0 | | | | | | | | |
| Cr | 87 | 134 | 97 | -60 | 106 | 251 | 65 | -14 | 230 | 0 | | | | | | | |
| C | 373 | 176 | -28 | -30 | 504 | -424 | -340 | 460 | -413 | 328 | 0 | | | | | | |
| Re | 38 | -113 | -182 | -238 | 126 | -74 | -295 | -99 | -120 | 92 | 810 | 0 | | | | | |
| Ru | 120 | -91 | -164 | -149 | 173 | -208 | -321 | -353 | -260 | 109 | 698 | -58 | 0 | | | | |
| Os | 119 | -68 | -143 | -147 | 212 | -126 | -296 | -165 | -206 | 92 | 944 | -75 | 116 | 0 | | | |
| Rh | -65 | -294 | -358 | -294 | -45 | -546 | -569 | -655 | -605 | 8 | 621 | -121 | 28 | 65 | 0 | | |
| Ir | -146 | -344 | -422 | -404 | -111 | -578 | -656 | -522 | -639 | -100 | 866 | -111 | -22 | 29 | -30 | 0 | |
| Si | -126 | -301 | -222 | -291 | 102 | -535 | -411 | 220 | -393 | -103 | 354 | -93 | -326 | -81 | -552 | -419 | 0 |

Note: The pairwise mixing enthalpy is calculated with respect to constituent elements in their stable states at T = 0K.

**Appendix B. $E_{hull}$ distribution for single-phase RCCAs**

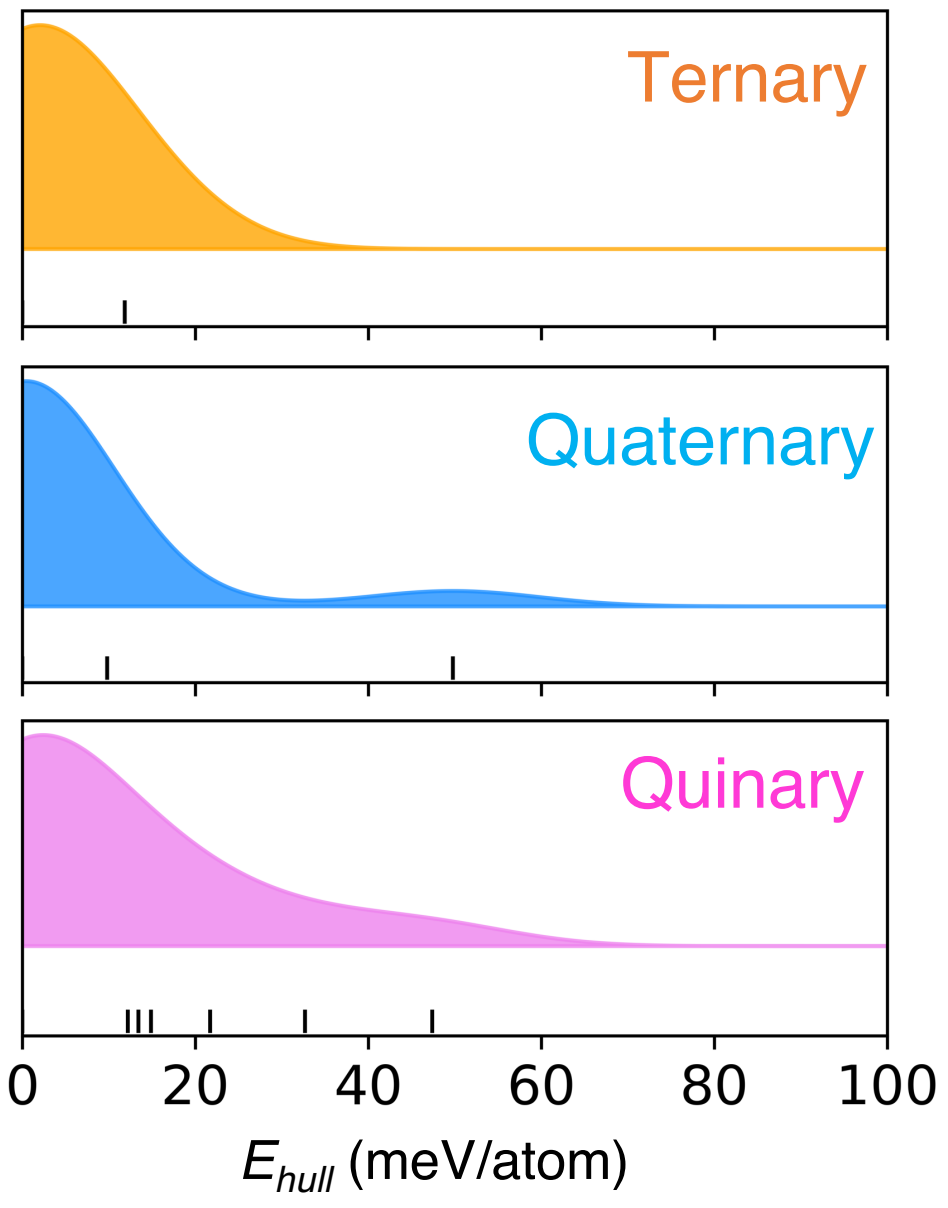

Fig. 6: Gaussian kernel density estimates of $E_{hull}$ distribution for single-phase ternary, quaternary and quinary RCCAs.

Table 1: DFT-calculated $E_{hull}$ of RCCAs that have been observed to form single-phase solid solution experimentally.

| Composition | $E_{hull}$(meV/atom) | Composition | $E_{hull}$(meV/atom) |
|---|---|---|---|
| MoNbTi | 0 | HfMoNbTaTi | 0 |
| NbTaTi | 0 | HfMoNbTaZr | 0 |
| NbTiZr | 0 | HfMoNbTiZr | 0 |
| MoNbTa | 12 | HfMoTaTiZr | 0 |
| AlMoNbTi | 0 | HfNbTaTiZr | 0 |
| HfNbTaTi | 0 | MoNbTaTiV | 0 |
| HfNbTaZr | 0 | MoNbTaTiW | 0 |
| HfNbTiZr | 0 | MoNbTaTiZr | 0 |
| HfTaTiZr | 0 | MoNbTaVW | 0 |
| MoNbTaTi | 0 | NbTaTiVW | 12 |
| MoNbTaW | 0 | CrMoNbTiW | 13 |
| MoNbTiV | 0 | MoNbTiVZr | 15 |
| MoNbTiZr | 0 | AlMoNbTiV | 22 |
| MoTaTiV | 0 | AlMoTaTiV | 33 |
| NbTaTiV | 0 | NbTaTiVZr | 47 |
| NbTaTiW | 0 | | |
| NbTaTiZr | 0 | | |
| NbTaVW | 0 | | |
| AlNbTiV | 10 | | |
| NbVZrTi | 50 | | |

**Appendix C. Phase prediction in NbTiZr-based RCCAs**

Table 2. Phases that are included in the convex hull analysis of NbTiZr alloy.

| composition | $\Delta H$(meV/atom) | $\Delta S$(meV/K) | Source |
|---|---|---|---|
| NbZrTi | 76 | -9.47E-02 | model |
| NbZr | 70 | -5.97E-02 | DFT |
| NbTi | 41 | -5.97E-02 | DFT |
| ZrTi | 90 | -5.97E-02 | DFT |
| Nb | 0 | 0 | mp-75 |
| Zr | 0 | 0 | mp-131 |
| Ti | 0 | 0 | mp-72 |

| |
|---|
| Solid solution |
| element |
| intermetallic |

| composition | $\Delta H$(meV/atom) | $\Delta S$(meV/K) | Source |
|---|---|---|---|
| $Ti_3Nb$ | 82 | 0 | mp-981232 |
| TiNb | 35 | 0 | mp-1216634 |
| $Ti_3Nb$ | 55 | 0 | mp-1217091 |
| $Ti_3Nb$ | 84 | 0 | mp-980945 |
| $Ti_3Nb$ | 78 | 0 | mp-1187514 |
| ZrNb | 111 | 0 | mp-1215202 |
| ZrTi | 88 | 0 | mp-1215236 |
| $ZrTi_2$ | 185 | 0 | mp-1080389 |
| $ZrTi_3$ | 31 | 0 | mp-1183046 |
| $ZrTi_2$ | 32 | 0 | mp-1008568 |
| ZrTi | 79 | 0 | mp-1215200 |
| ZrTiNb | 110 | 0 | mp-1215185 |

Table 3. Phases that are included in the convex hull analysis of NbTiZrV alloy.

| composition | $\Delta H$(meV/atom) | $\Delta S$(meV/K) | Source |
|---|---|---|---|
| NbVZrTi | 104 | -1.19E-01 | model |
| NbVZr | 149 | -9.47E-02 | model |
| NbVTi | 72 | -9.47E-02 | model |
| NbZrTi | 76 | -9.47E-02 | model |
| VZrTi | 140 | -9.47E-02 | model |
| NbV | 75 | -5.97E-02 | DFT |
| NbZr | 70 | -5.97E-02 | DFT |
| NbTi | 41 | -5.97E-02 | DFT |
| VZr | 195 | -5.97E-02 | DFT |
| VTi | 86 | -5.97E-02 | DFT |
| ZrTi | 90 | -5.97E-02 | DFT |
| Nb | 0 | 0 | mp-75 |
| V | 0 | 0 | mp-146 |
| Zr | 0 | 0 | mp-131 |
| Ti | 0 | 0 | mp-72 |

| composition | $\Delta H$(meV/atom) | $\Delta S$(meV/K) | Source |
|---|---|---|---|
| $NbV_2$ | -59 | 0 | [11] |
| $Ti_3Nb$ | 82 | 0 | mp-981232 |
| TiNb | 35 | 0 | mp-1216634 |
| $Ti_3Nb$ | 55 | 0 | mp-1217091 |
| $Ti_3Nb$ | 84 | 0 | mp-980945 |
| $Ti_3Nb$ | 78 | 0 | mp-1187514 |
| ZrNb | 111 | 0 | mp-1215202 |
| TiV | 118 | 0 | mp-1216646 |
| $Ti_4V$ | 121 | 0 | mp-1217117 |
| ZrTi | 88 | 0 | mp-1215236 |
| $ZrTi_2$ | 185 | 0 | mp-1080389 |
| $ZrTi_3$ | 31 | 0 | mp-1183046 |
| $ZrTi_2$ | 32 | 0 | mp-1008568 |
| ZrTi | 79 | 0 | mp-1215200 |
| $Zr_3V$ | 216 | 0 | mp-1188058 |
| $ZrV_2$ | 51 | 0 | mp-258 |
| ZrTiNb | 110 | 0 | mp-1215185 |

Table 4. Phases that are included in the convex hull analysis of NbTiZrVMo alloy.

| composition | $\Delta H$(meV/atom) | $\Delta S$(meV/K) | Source |
|---|---|---|---|
| MoNbTiVZr | 32 | 1.39E-01 | model |
| MoNbTiV | -26 | 1.19E-01 | model |
| MoNbTiZr | 9 | 1.19E-01 | model |
| MoTiVZr | 27 | 1.19E-01 | model |
| NbTiVZr | 103 | 1.19E-01 | model |
| MoNbVZr | 36 | 1.19E-01 | model |
| MoNbTi | -53 | 9.47E-02 | model |
| MoNbV | -49 | 9.47E-02 | model |
| MoNbZr | 14 | 9.47E-02 | model |
| MoTiV | -58 | 9.47E-02 | model |
| MoTiZr | 4 | 9.47E-02 | model |
| MoVZr | 38 | 9.47E-02 | model |
| NbTiV | 68 | 9.47E-02 | model |
| NbTiZr | 66 | 9.47E-02 | model |
| NbVZr | 122 | 9.47E-02 | model |
| TiVZr | 110 | 9.47E-02 | model |
| MoNb | -73 | 5.97E-02 | DFT |
| MoTi | -81 | 5.97E-02 | DFT |
| MoV | -102 | 5.97E-02 | DFT |
| MoZr | 42 | 5.97E-02 | DFT |
| NbTi | 35 | 5.97E-02 | DFT |
| NbV | 65 | 5.97E-02 | DFT |
| NbZr | 64 | 5.97E-02 | DFT |
| TiV | 52 | 5.97E-02 | DFT |
| TiZr | 49 | 5.97E-02 | DFT |
| VZr | 146 | 5.97E-02 | DFT |
| Mo | 0 | 0 | mp-129 |
| Nb | 0 | 0 | mp-75 |
| Ti | 0 | 0 | mp-131 |
| V | 0 | 0 | mp-146 |
| Zr | 0 | 0 | mp-72 |

| composition | $\Delta H$(meV/atom) | $\Delta S$(meV/K) | Source |
|---|---|---|---|
| $NbV_2$ | -59 | 0 | [11] |
| NbMo | -74 | 0 | mp-1220327 |
| TiMo | 701 | 0 | mp-998968 |
| $TiMo_2$ | -88 | 0 | mp-1216675 |
| $TiMo_3$ | -147 | 0 | mp-1017983 |
| VMo | 343 | 0 | mp-995205 |
| $V_3Mo$ | -68 | 0 | mp-972071 |
| $Zr_4Mo$ | 910 | 0 | mp-1207454 |
| $ZrMo_3$ | 228 | 0 | mp-30790 |
| $ZrMo_2$ | 58 | 0 | mp-1215231 |
| ZrMo | 105 | 0 | mp-1215206 |
| $ZrMo_2$ | -139 | 0 | mp-2049 |
| $Ti_3Nb$ | 82 | 0 | mp-981232 |
| TiNb | 35 | 0 | mp-1216634 |
| $Ti_3Nb$ | 55 | 0 | mp-1217091 |
| $Ti_3Nb$ | 84 | 0 | mp-980945 |
| $Ti_3Nb$ | 78 | 0 | mp-1187514 |
| ZrNb | 111 | 0 | mp-1215202 |
| TiV | 118 | 0 | mp-1216646 |
| $Ti_4V$ | 121 | 0 | mp-1217117 |
| ZrTi | 88 | 0 | mp-1215236 |
| $ZrTi_2$ | 185 | 0 | mp-1080389 |
| $ZrTi_3$ | 31 | 0 | mp-1183046 |
| $ZrTi_2$ | 32 | 0 | mp-1008568 |
| ZrTi | 79 | 0 | mp-1215200 |
| $Zr_3V$ | 216 | 0 | mp-1188058 |
| $ZrV_2$ | 51 | 0 | mp-258 |
| $Ti_2NbMo$ | 3986 | 0 | mp-1096197 |
| $TiNb_2Mo$ | 4257 | 0 | mp-1097379 |
| $NbV_2Mo$ | 4262 | 0 | mp-1095929 |
| $NbVMo_2$ | 4455 | 0 | mp-1096452 |
| $Nb_2VMo$ | 4316 | 0 | mp-1095742 |
| $Ti_2VMo$ | 3772 | 0 | mp-1096187 |
| $TiV_2Mo$ | 4196 | 0 | mp-1095882 |
| $TiVMo_2$ | 4154 | 0 | mp-1096250 |
| $ZrTiMo_4$ | -60 | 0 | mp-1215177 |
| ZrVMo | -115 | 0 | mp-1215168 |
| ZrTiNb | 110 | 0 | mp-1215185 |

Table 5. Phases that are included in the convex hull analysis of NbTiZrVTa alloy.

| composition | Δ*H*(meV/atom) | Δ*S*(meV/K) | Source |
|---|---|---|---|
| NbTaTiVZr | 102 | 1.39E-01 | model |
| NbTaTiV | 67 | 1.19E-01 | model |
| NbTaTiZr | 79 | 1.19E-01 | model |
| NbTiVZr | 103 | 1.19E-01 | model |
| TaTiVZr | 118 | 1.19E-01 | model |
| NbTaVZr | 110 | 1.19E-01 | model |
| NbTaTi | 43 | 9.47E-02 | model |
| NbTaV | 54 | 9.47E-02 | model |
| NbTaZr | 77 | 9.47E-02 | model |
| NbTiV | 68 | 9.47E-02 | model |
| NbTiZr | 66 | 9.47E-02 | model |
| NbVZr | 122 | 9.47E-02 | model |
| TaTiV | 75 | 9.47E-02 | model |
| TaTiZr | 97 | 9.47E-02 | model |
| TaVZr | 138 | 9.47E-02 | model |
| TiVZr | 110 | 9.47E-02 | model |
| NbTa | 0 | 5.97E-02 | DFT |
| NbTi | 35 | 5.97E-02 | DFT |
| NbV | 65 | 5.97E-02 | DFT |
| NbZr | 64 | 5.97E-02 | DFT |
| TaTi | 61 | 5.97E-02 | DFT |
| TaV | 55 | 5.97E-02 | DFT |
| TaZr | 109 | 5.97E-02 | DFT |
| TiV | 52 | 5.97E-02 | DFT |
| TiZr | 49 | 5.97E-02 | DFT |
| VZr | 146 | 5.97E-02 | DFT |
| Nb | 0 | 0 | mp-75 |
| Ta | 0 | 0 | mp-50 |
| Ti | 0 | 0 | mp-72 |
| V | 0 | 0 | mp-146 |
| Zr | 0 | 0 | mp-131 |
| $NbV_2$ | -59 | 0 | [11] |
| TaNb | 8 | 0 | mp-1217892 |
| $Ti_3Nb$ | 82 | 0 | mp-981232 |
| TiNb | 35 | 0 | mp-1216634 |
| $Ti_3Nb$ | 55 | 0 | mp-1217091 |
| $Ti_3Nb$ | 84 | 0 | mp-980945 |
| $Ti_3Nb$ | 78 | 0 | mp-1187514 |
| ZrNb | 111 | 0 | mp-1215202 |
| $TaTi_3$ | 97 | 0 | mp-1187256 |
| TaTi | 60 | 0 | mp-1217887 |
| $TaTi_3$ | 97 | 0 | mp-1187253 |
| $TaTi_3$ | 95 | 0 | mp-1187250 |
| $TaV_2$ | -103 | 0 | mp-567276 |
| TaV | 86 | 0 | mp-1217812 |
| $Zr_3Ta$ | 153 | 0 | mp-1188053 |
| $Zr_3Ta$ | 167 | 0 | mp-1188024 |
| TiV | 118 | 0 | mp-1216646 |
| $Ti_4V$ | 121 | 0 | mp-1217117 |
| ZrTi | 88 | 0 | mp-1215236 |
| $ZrTi_2$ | 185 | 0 | mp-1080389 |
| $ZrTi_3$ | 31 | 0 | mp-1183046 |
| $ZrTi_2$ | 32 | 0 | mp-1008568 |
| ZrTi | 79 | 0 | mp-1215200 |
| $Zr_3V$ | 216 | 0 | mp-1188058 |
| ZrV2 | 51 | 0 | mp-258 |
| $TaTiNb_2$ | 4424 | 0 | mp-1097329 |
| TaNbV | 45 | 0 | mp-1217905 |
| ZrTiNb | 110 | 0 | mp-1215185 |

Table 6. Phases that are included in the convex hull analysis of NbTiZrVCr alloy.

| composition | Δ*H*(meV/atom) | Δ*S*(meV/K) | Source |
|---|---|---|---|
| CrNbTiVZr | 32 | 1.39E-01 | model |
| CrNbTiV | -26 | 1.19E-01 | model |
| CrNbTiZr | 9 | 1.19E-01 | model |
| CrTiVZr | 27 | 1.19E-01 | model |
| NbTiVZr | 103 | 1.19E-01 | model |
| CrNbVZr | 36 | 1.19E-01 | model |
| CrNbTi | -53 | 9.47E-02 | model |
| CrNbV | -49 | 9.47E-02 | model |
| CrNbZr | 14 | 9.47E-02 | model |
| CrTiV | -58 | 9.47E-02 | model |
| CrTiZr | 4 | 9.47E-02 | model |
| CrVZr | 38 | 9.47E-02 | model |
| NbTiV | 68 | 9.47E-02 | model |
| NbTiZr | 66 | 9.47E-02 | model |
| NbVZr | 122 | 9.47E-02 | model |
| TiVZr | 110 | 9.47E-02 | model |
| CrNb | -73 | 5.97E-02 | DFT |
| CrTi | -81 | 5.97E-02 | DFT |
| CrV | -102 | 5.97E-02 | DFT |
| CrZr | 42 | 5.97E-02 | DFT |
| NbTi | 35 | 5.97E-02 | DFT |
| NbV | 65 | 5.97E-02 | DFT |
| NbZr | 64 | 5.97E-02 | DFT |
| TiV | 52 | 5.97E-02 | DFT |
| TiZr | 49 | 5.97E-02 | DFT |
| VZr | 146 | 5.97E-02 | DFT |
| Cr | 0 | 0 | mp-90 |
| Nb | 0 | 0 | mp-75 |
| Ti | 0 | 0 | mp-72 |
| V | 0 | 0 | mp-146 |
| Zr | 0 | 0 | mp-131 |

| composition | Δ*H*(meV/atom) | Δ*S*(meV/K) | Source |
|---|---|---|---|
| NbV2 | -59 | 0 | [11] |
| Nb3Cr | -74 | 0 | mp-999446 |
| NbCr3 | 701 | 0 | mp-999392 |
| NbCr2 | -88 | 0 | mp-1220609 |
| NbCr3 | -147 | 0 | mp-999393 |
| Nb3Cr | 343 | 0 | mp-999441 |
| NbCr2 | -68 | 0 | mp-548 |
| NbCr2 | 910 | 0 | mp-1095643 |
| NbCr2 | 228 | 0 | mp-1191777 |
| Nb2Cr | 58 | 0 | mp-1077258 |
| NbCr3 | 105 | 0 | mp-999390 |
| Nb3Cr | -139 | 0 | mp-999439 |
| TiCr2 | 82 | 0 | mp-1425 |
| TiCr2 | 35 | 0 | mp-568636 |
| TiCr2 | 55 | 0 | mp-1589 |
| Ti4Cr | 84 | 0 | mp-1217156 |
| VCr3 | 78 | 0 | mp-1187696 |
| V3Cr | 111 | 0 | mp-1187695 |
| VCr | 118 | 0 | mp-1216394 |
| ZrCr2 | 121 | 0 | mp-570608 |
| ZrCr2 | 88 | 0 | mp-903 |
| ZrCr2 | 185 | 0 | mp-1919 |
| Ti3Nb | 31 | 0 | mp-981232 |
| TiNb | 32 | 0 | mp-1216634 |
| Ti3Nb | 79 | 0 | mp-1217091 |
| Ti3Nb | 216 | 0 | mp-980945 |
| Ti3Nb | 51 | 0 | mp-1187514 |
| ZrNb | 3986 | 0 | mp-1215202 |
| TiV | 4257 | 0 | mp-1216646 |
| Ti4V | 4262 | 0 | mp-1217117 |
| ZrTi | 4455 | 0 | mp-1215236 |
| ZrTi2 | 4316 | 0 | mp-1080389 |
| ZrTi3 | 3772 | 0 | mp-1183046 |
| ZrTi2 | 4196 | 0 | mp-1008568 |
| ZrTi | 4154 | 0 | mp-1215200 |
| Zr3V | -60 | 0 | mp-1188058 |
| ZrV2 | -115 | 0 | mp-258 |
| TiNbCr4 | 110 | 0 | mp-1216666 |
| NbVCr | -0.09 | 0 | mp-1220374 |
| ZrNbCr4 | -0.04 | 0 | mp-1215217 |
| ZrTiCr4 | -0.04 | 0 | mp-1215221 |
| ZrTiCr4 | -0.05 | 0 | mp-1215179 |
| ZrVCr | -0.02 | 0 | mp-1215170 |
| ZrTiNb | 0.11 | 0 | mp-1215185 |

## Appendix D. Single-phase BCC RCCAs candidates

### Quaternary alloys

| 1000 K (meV/atom) | | | | | | | | | |
|---|---|---|---|---|---|---|---|---|---|
| Composition | $E_{hull}$ | Composition | $E_{hull}$ | Composition | $E_{hull}$ | Composition | $E_{hull}$ | Composition | $E_{hull}$ |
| HfNbVZr | 3 | CrMoNbOs | 11 | CrTaTiZr | 15 | HfTaVZr | 17 | NbReVZr | 19 |
| CrMoReW | 3 | MoNbOsRu | 11 | CrHfTaTi | 15 | CrMoReTa | 17 | CrMoRhRu | 19 |
| HfNbVW | 5 | CrNbTaTi | 11 | CrMoOsV | 15 | CrTaTiW | 17 | CrHfTaV | 19 |
| HfMoNbV | 7 | CrMoRhW | 11 | HfMoNbW | 16 | HfNbReV | 17 | CrHfNbV | 19 |
| CrTaTiV | 8 | CrMoNbRe | 11 | MoNbWZr | 16 | CrNbReW | 18 | CrRuVW | 20 |
| HfTaTiV | 8 | CrMoReRu | 12 | CrMoRuV | 16 | CrMoNbRu | 18 | CrHfTiZr | 20 |
| NbVWZr | 8 | CrMoOsW | 12 | HfMoTiW | 16 | HfMoWZr | 18 | MoRhRuW | 20 |
| NbReTaZr | 8 | MoOsReRu | 13 | CrNbOsW | 16 | CrNbOsRu | 18 | MoOsRuW | 20 |
| AlCrVW | 9 | CrHfNbTi | 14 | CrMoOsTa | 16 | ReTaVZr | 18 | CrMoOsRh | 20 |
| CrReRhRu | 9 | CrReRhW | 14 | HfReTaV | 16 | CrReRuW | 18 | HfMoVW | 20 |
| MoNbVZr | 10 | CrMoOsRe | 14 | MoTiWZr | 17 | CrMoReV | 18 | HfMoTaW | 20 |
| HfNbReTa | 10 | CrNbTiW | 15 | MoOsRuV | 17 | CrMoTaTi | 19 | AlNbTiV | 20 |
| CrOsReRu | 10 | | | | | | | | |

| 2000 K (meV/atom) | | | | | | | | | |
|---|---|---|---|---|---|---|---|---|---|
| Composition | $E_{hull}$ | Composition | $E_{hull}$ | Composition | $E_{hull}$ | Composition | $E_{hull}$ | Composition | $E_{hull}$ |
| CrMoOsRu | 0 | HfMoNbTi | 9 | CrNbTiZr | 13 | CrOsRhW | 16 | CrRuVW | 18 |
| CrOsRhRu | 0 | OsReRhRu | 9 | MoNbWZr | 13 | CrMoRuV | 16 | CrNbOsRu | 18 |
| HfNbTiV | 0 | CrMoTiW | 9 | CrMoNbOs | 14 | HfMoReZr | 16 | CrMoNbTi | 18 |
| NbTiVZr | 0 | MoOsRuV | 9 | CrMoOsRh | 14 | HfMoNbV | 16 | OsRhRuW | 18 |
| IrOsRhRu | 2 | MoNbTaTi | 9 | CrMoNbW | 14 | CrTiWZr | 16 | MoNbVZr | 18 |
| HfNbTaV | 2 | HfNbTaTi | 10 | CrMoVW | 14 | CrNbTiW | 16 | MoOsRhW | 18 |
| NbTaVZr | 3 | CrOsReRu | 10 | MoOsRhRu | 14 | MoOsRuTa | 16 | CrNbTaW | 18 |
| MoNbOsRu | 3 | MoNbTiV | 10 | CrReRhW | 14 | CrMoOsW | 16 | MoOsReV | 19 |
| TaTiVZr | 4 | MoOsReRu | 10 | NbTaWZr | 14 | IrReRhRu | 16 | CrMoReSi | 19 |
| NbTaTiV | 4 | CrMoRuW | 10 | CrTaTiZr | 14 | CrReRhRu | 16 | CrMoNbRu | 19 |
| CrOsRuW | 4 | NbTiWZr | 11 | CrMoOsV | 14 | MoReRuV | 16 | CrMoOsRe | 19 |
| HfNbTiZr | 5 | IrOsReRh | 11 | HfMoTiW | 15 | NbOsRuW | 17 | MoNbTiW | 19 |
| HfTaTiV | 5 | CrMoReRh | 11 | CrHfTaTi | 15 | IrOsReRu | 17 | MoOsRuW | 19 |
| CrMoReW | 5 | CrNbTaTi | 11 | MoNbTaZr | 15 | CrTaTiV | 17 | CrOsVW | 19 |
| CrMoRhW | 6 | NbTaVW | 11 | MoRhRuW | 15 | CrMoNbRe | 17 | HfTaVZr | 19 |
| HfTaTiZr | 6 | HfNbTiW | 11 | MoNbTaV | 15 | CrMoReRu | 17 | NbRuTaV | 19 |
| TaTiVW | 7 | HfTiVZr | 11 | HfNbReTa | 15 | HfReTaV | 17 | HfMoTiV | 20 |
| NbTaTiZr | 8 | CrIrOsRu | 11 | HfNbVW | 15 | CrNbTaZr | 17 | HfNbTaZr | 20 |
| NbTaTiW | 8 | HfNbVZr | 12 | CrMoWZr | 15 | NbVWZr | 17 | MoTiVZr | 20 |
| CrRhRuW | 8 | CrOsRuV | 12 | TaVWZr | 15 | NbReTaZr | 17 | MoNbTaW | 20 |
| MoTaTiV | 8 | HfNbTaW | 12 | HfMoWZr | 15 | CrIrRhRu | 17 | AlCrTaW | 20 |
| AlCrVW | 8 | HfMoTaV | 12 | MoTaVZr | 15 | CrNbOsW | 18 | CrMoTiZr | 20 |
| NbTiVW | 8 | HfTaVW | 12 | CrHfNbTi | 15 | AlNbTaV | 18 | CrOsReRh | 20 |
| CrMoRhRu | 8 | CrIrOsRh | 13 | MoTiWZr | 15 | AlMoVW | 18 | CrNbRuW | 20 |
| MoNbTiZr | 8 | HfMoNbTa | 13 | HfNbReZr | 15 | CrTaTiW | 18 | HfTiVW | 20 |
| AlCrMoW | 8 | HfMoNbW | 13 | CrMoTaW | 15 | | | | |

## Quinary alloys

| 1000 K (meV/atom) | | | | | | | | | |
|---|---|---|---|---|---|---|---|---|---|
| Composition | $E_{hull}$ | Composition | $E_{hull}$ | Composition | $E_{hull}$ | Composition | $E_{hull}$ | Composition | $E_{hull}$ |
| HfNbTiVZr | 9 | HfNbTiVW | 14 | HfMoNbTaTi | 16 | HfMoNbTiZr | 17 | MoNbTiVW | 20 |
| HfNbTaTiZr | 12 | NbTaTiWZr | 15 | HfMoNbTiV | 16 | HfTaTiVZr | 18 | CrIrOsRhRu | 20 |
| NbTaTiVW | 13 | HfNbTaTiW | 15 | MoNbTaTiZr | 16 | CrOsReRhRu | 19 | MoNbTaTiW | 20 |
| NbTaTiVZr | 13 | NbTiVWZr | 15 | HfNbTiWZr | 17 | HfNbTaVW | 19 | NbTaVWZr | 20 |
| HfNbTaTiV | 14 | MoNbTaTiV | 15 | MoNbTiVZr | 17 | IrOsReRhRu | 19 | | |

| 2000 K (meV/atom) | | | | | | | | | |
|---|---|---|---|---|---|---|---|---|---|
| Composition | $E_{hull}$ | Composition | $E_{hull}$ | Composition | $E_{hull}$ | Composition | $E_{hull}$ | Composition | $E_{hull}$ |
| NbTaTiVZr | 7 | CrOsRhRuW | 15 | CrIrOsRhRu | 17 | HfNbTaVZr | 18 | CrMoNbTiZr | 19 |
| HfNbTaTiV | 8 | CrNbTaTiZr | 15 | CrHfNbTiZr | 17 | NbTaTiWZr | 18 | HfMoNbTaTi | 19 |
| CrMoRhRuW | 11 | HfNbTaTiZr | 15 | MoOsRhRuW | 17 | HfMoNbTiW | 18 | HfMoNbTiV | 19 |
| HfNbTiVZr | 12 | HfNbTaVW | 15 | CrOsRuVW | 17 | NbTiVWZr | 18 | MoNbTaTiZr | 19 |
| NbTaTiVW | 12 | CrMoNbOsRu | 15 | HfMoNbTaV | 17 | CrOsReRuW | 18 | HfMoTaTiV | 19 |
| CrMoOsReRu | 14 | CrMoOsRuV | 15 | IrOsReRhRu | 18 | CrNbTiWZr | 18 | MoNbTiWZr | 20 |
| CrMoOsRhRu | 14 | CrMoOsRhW | 16 | HfNbTiVW | 18 | HfTaTiVW | 19 | HfNbTiWZr | 20 |
| HfTaTiVZr | 14 | NbTaVWZr | 16 | TaTiVWZr | 18 | CrNbTaTiW | 19 | MoNbTiVZr | 20 |
| MoNbTaTiV | 15 | CrOsReRhRu | 16 | HfNbTaTiW | 18 | MoTaTiVZr | 19 | HfMoNbTiZr | 20 |
| CrMoOsRuW | 15 | CrNbOsRuW | 17 | CrHfNbTaTi | 18 | MoNbTaVZr | 19 | CrHfTaTiZr | 20 |

## Senary alloys

| 1000 K (meV/atom) | | | | | | | | | |
|---|---|---|---|---|---|---|---|---|---|
| Composition | $E_{hull}$ | Composition | $E_{hull}$ | Composition | $E_{hull}$ | Composition | $E_{hull}$ | Composition | $E_{hull}$ |
| HfNbTaTiVZr | 17 | HfNbTaTiVW | 17 | HfMoNbTiVZr | 19 | HfNbTiVWZr | 20 | MoNbTiVW | 20 |
| NbTaTiVWZr | 17 | HfMoNbTaTiV | 19 | MoNbTaTiVZr | 19 | | | | |

| 2000 K (meV/atom) | | | | | | | | | |
|---|---|---|---|---|---|---|---|---|---|
| Composition | $E_{hull}$ | Composition | $E_{hull}$ | Composition | $E_{hull}$ | Composition | $E_{hull}$ | Composition | $E_{hull}$ |
| HfNbTaTiVZr | 16 | CrMoOsRhRuW | 17 | CrHfNbTaTiZr | 20 | HfMoNbTaTiV | 20 | NbTaTiVWZr | 20 |

**Appendix E. Comparison of our model with a similar model reported by Bokas et al. [Ref. 23]**

As discussed in the manuscript, our approach of using pairwise mixing enthalpies to estimate the mixing enthalpy of HEAs is similar to that proposed by Bokas et al.[23]. For the 14 elements that we have in common (Al, Ti, V, Cr, Zr, Nb, Mo, Hf, Ta, W, Re, Os, Rh, and Ir), we compare the mixing enthalpy of their ternary, quaternary and quinary alloys using both the models. The results are shown in Fig. 7. Overall, we find a good agreement between both the models with an MAE of 0.035 eV/atom between the two models. However, alloys containing Hf-Ti, Re-Mo, and Os-Mo pairs show large differences. For instance, the model by Bokas *et al.*, gives lower $\Delta H_{mix}$ predictions on HfTiV and ReMoOs compared to ours. This difference can be tracked back to the different pairwise mixing enthalpy of Hf-Ti, Re-Mo, and Os-Mo in the two models. We obtain mixing enthalpy of Hf-Ti, Re-Mo, and Os-Mo as 0.04, 0.04, and 0.11 eV/atom, respectively. For the respective pairs, Bokas et al. report values of –0.36, –0.04, and –0.21 eV/atom. These differences may arise from the different values of the cutoff energy and *k*-points used for the DFT calculations in the two studies; we have used tighter convergence settings.

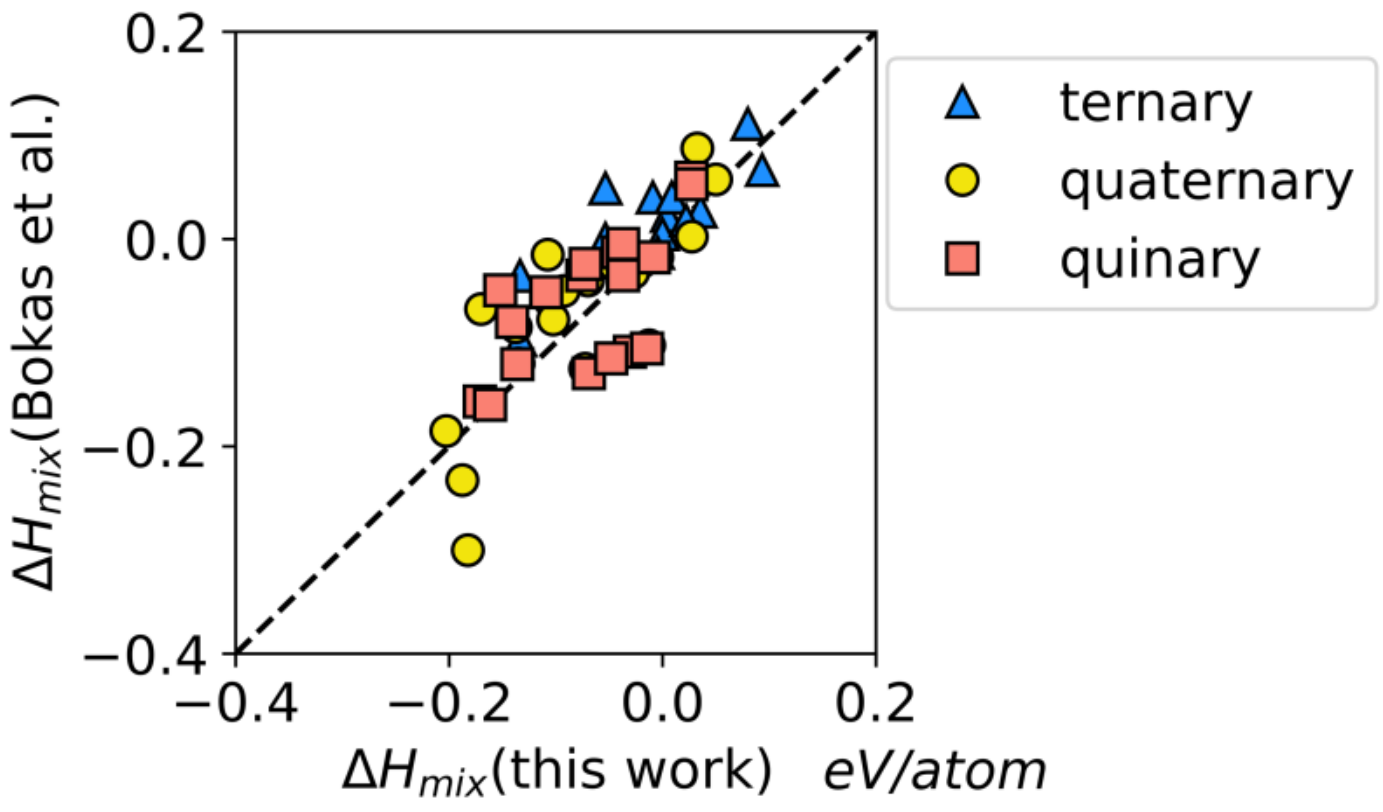

Fig. 7: A comparison of predicted $\Delta H_{mix}$ using our model and the model by Bokas et al. (Ref [23]) for ternary, quaternary and quinary alloys.